# Comment on: "A significant enhancement in visible-light photodetection properties of chemical spray pyrolysis fabricated CdS thin films by novel Eu doping concentrations" [Sens. Actuator. A Phys. 301 (2020) 111749].

Ramezan Sahebi [a]

[a] Thin films laboratory, Department of physics, Faculty of Science, Ferdowsi University of Mashhad, Mashhad, Iran

Email addresses: ramezan.sahebi@mail.um.ac.ir

**Abstract**

In a recent paper [Sens. Actuator. A Phys. 301 (2020) 111749] Shkir et al. have studied the structural, optical and photo-electrical properties of the CdS thin films as a function of Eu doping concentration. The authors have used a wrong equation to calculate the refractive index and obtained values are incorrect. Consequently, the optical properties obtained based on the refractive index such as real and imaginary parts of dielectric constant, dielectric loss, optical conductivity, nonlinear refractive index ($n_2$), linear and third-order non-linear optical susceptibility are wrong.

In the commented paper, Shkir et al [1] have studied the physical properties of CdS thin films as a function of Eu doping. They have measured the transmittance, reflectance and absorbance spectra of the samples and used these data to study the linear and non-linear optical properties of the specimens. The refractive index of the samples has been evaluated by using a wrong equation (equation (7) in the commented paper):

$$n^2 = N + (N^2 - n_a^2 n_s^2)^{\frac{1}{2}} \ where \ N = \frac{n_a^2 + n_s^2}{2} + 2n_a n_s T \qquad (1)$$

where n is the refractive index, T is the transmittance, $n_a$ and $n_s$ are the refractive index of the air and substrate, respectively. It must be mentioned that this equation is not used in the presented reference by the authors (reference (66) in the commented paper). The applied method to calculate the refractive index of thin films using transmittance spectra was proposed by Swanepoel [2]. According to this method, if the interference fringes are observed in the transmittance spectra, the refractive index of the thin films can be calculated based on the maximum $T_M$ and minimum $T_m$ transmittance envelope curves obtained by parabolic interpolation to the experimentally determined positions of peaks and valleys. The refractive index must be calculated by the following equation:

$$n^2 = N + (N^2 - n_a^2 n_s^2)^{\frac{1}{2}} \ where \ N = \frac{n_a^2 + n_s^2}{2} + \frac{2n_a n_s (T_M - T_m)}{T_M T_m} \qquad (2)$$

As shown, the calculated values of the refractive index in the commented paper are incorrect. Consequently, the optical properties obtained based on the refractive index such as real and imaginary parts of dielectric constant ($\varepsilon_r = n^2 - k^2, \varepsilon_i = 2nk$), dielectric loss ($\tan(\frac{\varepsilon_i}{\varepsilon_r})$) and optical conductivity ($\sigma = \frac{\alpha n c}{4\pi}$) are wrong. Also, some of the non-linear optical properties of the samples are calculated in the commented paper by using the following equations (these equations are listed in the reference (72) of the commented paper):

$$\chi^{(1)} = \frac{n_0^2 - 1}{4\pi} \text{ and } \chi^{(3)} = 1.7 \times 10^{-10} \left(\chi^{(1)}\right)^4 \text{ and } n_2 = \frac{12\pi\chi^{(3)}}{n_0} \tag{3}$$

where $n_0$, $n_2$ are the linear and non-linear parts of the refractive index ($n = n_0(\lambda) + n_2(E^2)$). $\chi^{(1)}$ and $\chi^{(3)}$ are the linear and third-order non-linear optical susceptibility. As mentioned in this comment, the linear part on the refractive index is calculated using a wrong equation and consequently, the obtained values of the non-linear optical parameters are incorrect.

It must be noted that the refractive index is a complex parameter (n$^*$=n+ik). The imaginary part, k, is known as the extinction coefficient [3]. In the commented paper this parameter is expressed using $k = \frac{\alpha\lambda}{4\pi}$ equation and called as absorption coefficient. The equation is correct and the authors just made a mistake in writing the name of this parameter.

It is worthwhile to mention that using the Swanepoel method to calculate the refractive index is not an ideal method for the commented paper. This method is suitable to estimate the refractive index for the researchers who are not equipped with the spectrometer capable of measuring the reflectance spectra and just can to measure the transmittance spectra of the specimens. In the process of obtaining the refractive index equation in this model, some simplifications have been made. For example, the extinction coefficient of the thin film is neglected in this process [2]. The authors of the commented paper have measured the transmittance and reflectance spectra of their samples and they can find more accurate and high- precise values of the refractive index by using the following relations [4]:

$$R = \frac{(n-1)^2 + k^2}{(n+1)^2 + k^2} \Rightarrow n = \frac{1+R}{1-R} + \sqrt{\frac{4R}{(1-R)^2} - k^2} \tag{4}$$